\documentclass[aps,prb,twocolumn,superscriptaddress]{revtex4-1}
\usepackage{amsmath, amssymb}
\usepackage{graphicx}
\usepackage{hyperref}
\usepackage{color}
\usepackage{dcolumn} %align on decimal

\newcommand{\commutator}[2]{\left[#1,\: #2\right]}
\renewcommand{\vec}[1]{{\bf #1}}

\begin{document}
\title{Transport and particle-hole asymmetry in graphene on boron nitride}
\author{Ashley M. DaSilva}
\affiliation{Department of Physics, The University of Texas at Austin, Austin, Texas 78712-1192, USA}
\author{Jeil Jung}
\affiliation{Graphene Research Centre and Department of Physics, National University of Singapore, 2 Science Drive 3, 117551, Singapore}
\affiliation{Department of Physics, University of Seoul, Seoul, 130-743, Korea}
\author{Shaffique Adam}
\affiliation{Graphene Research Centre and Department of Physics, National University of Singapore, 2 Science Drive 3, 117551, Singapore}
\affiliation{Yale-NUS College, 6 College Avenue East, 138614, Singapore}
\author{Allan H. MacDonald}
\affiliation{Department of Physics, The University of Texas at Austin, Austin, Texas 78712-1192, USA}

\begin{abstract}
All local electronic properties of graphene on a hexagonal boron nitride (hBN) substrate
exhibit spatial moir{\'e} patterns related to lattice constant  and orientation differences  
between shared triangular Bravais lattices.
We apply a previously derived effective Hamiltonian for the $\pi$-bands of graphene on hBN to 
address the carrier-dependence of transport properties, concentrating
on the conductivity features at four electrons and four holes per unit cell.
These transport features measure the strength of 
Bragg scattering of $\pi$-electrons off the moir{\'e} pattern, and 
exhibit a striking particle-hole asymmetry that we trace to specific features of the effective 
Hamiltonian that we interpret physically.   
\end{abstract}

\maketitle

\section{Introduction}
Boron nitride is a popular substrate for high quality graphene devices because it is atomically smooth, has low 
chemical reactivity, and is typically relatively free of defects.\cite{dean_boron_2010,xue_scanning_2011,decker_local_2011,das_sarma_conductivity_2011,wong_characterization_2014}
These properties make it possible to achieve high graphene mobilities on hBN substrates.
Hexagonal boron nitride (hBN) is a wide band gap semiconductor\cite{watanabe_direct-bandgap_2004} 
with weakly coupled honeycomb lattice layers that are identical to graphene in structure, apart from a 
lattice constant difference of about two percent larger and possible differences in orientation.
If the graphene and hBN lattices were perfectly matched,
the graphene $\pi$ electrons would inherent hBN's broken inversion symmetry and develop a finite energy
gap at neutrality.\cite{giovannetti_substrate-induced_2007,jung_ab_2014}  
When exfoliated graphene is placed on a hBN surface, however, 
the two lattices do not align\cite{ortix_graphene_2012,xue_scanning_2011} 
and the electronic structure is more complex.\cite{wallbank_moire_2014,jung_origin_2015}  

In the most interesting case of nearly aligned layers,
the two similar lattice periodicities lead to  
long-period moir{\'e} patterns\cite{xue_scanning_2011,decker_local_2011,yankowitz_emergence_2012}  and 
to low-energy electronic properties that are insensitive to commensurability at an atomic level.
Experiments show that in samples of this type a gap nevertheless appears at the Fermi level of a neutral sheet,
and that its value is enhanced by electron-electron interactions and influenced
by the strain patterns induced in the graphene sheet by the 
lattice constant mismatch.\cite{hunt_massive_2013,ponomarenko_cloning_2013,woods_commensurate-incommensurate_2014} 
At zero rotation angle (perfect alignment), the moir{\'e} wavelength is around 15 nm.  In this case, secondary gapped Dirac points 
appear\cite{park_new_2008,park_new_2008,yankowitz_emergence_2012,
ortix_graphene_2012,wallbank_generic_2013,mucha-kruczynski_heterostructures_2013,jung_origin_2015}
at energies $\sim \pm$ 150 meV from the principal Dirac point, corresponding to the Fermi level of samples with $\pm 4$ electrons per moir{\'e} period, corresponding to one full conduction band or one empty valance band, respectively. 
These electronic structure features are conveniently probed by studying the
corresponding features in the carrier dependence of the {\em dc} transport properties of the  
graphene sheet. In this article, we present a theory of the transport properties of graphene on hexagonal boron nitride.
We focus on the $\pm 4$ electron-per-period features associated with the secondary Dirac points,
and in particular on the substantial particle-hole asymmetry which appears in all electronic properties, including 
the density of states.\cite{yankowitz_emergence_2012,wallbank_generic_2013,mucha-kruczynski_heterostructures_2013,martinez-gordillo_transport_2014} 
We trace the asymmetry to the influence of the moir{\'e} pattern on 
inter-sublattice hopping terms in the graphene sheet Hamiltonian, 
and in particular to differences between the hopping amplitudes to nearest neighbor sites 
at carbon-above-boron and carbon-above-nitrogen positions.  

Our paper is organized as follows. First in Section II we briefly summarize the moir{\'e} band Hamiltonian we use as the basis for our transport theory. 
In Section III we describe its band structure and discuss its density-of-states, carefully analyzing the origin of its particle/hole asymmetry. 
In Section IV we detail the transport theory we employ and in Section V we summarize its predictions 
for graphene on hBN. Finally, in Section VI we discuss and summarize our findings.  

\section{Moir{\'e} band model}

The moir{\'e} band Hamiltonian we employ for the graphene sheet $\pi$-band electrons accounts for the influence of the 
substrate by adding a local sub-lattice dependent term to the $\vec{k} \cdot \vec{p}$ Dirac model of an isolated graphene sheet.  
The substrate interaction term has the same periodicity as the moir{\'e} pattern and, because it is periodic,
can be analyzed using Bloch's theorem.  The moir{\'e} band Hamiltonian yields a non-trivial band structure.  
It does not account for those features of the full Hamiltonian associated with atomic-scale commensurability and 
is accurate only for moir{\'e} pattern periods that greatly exceed the graphene sheet lattice constant.  
This however is the interesting case, because the influence of the substrate is weak for 
short moir{\'e} periods.   The latter property can be traced\cite{bistritzer_moire_2011} in part 
to the property that the distance between layers is substantially larger than the distance between atoms within a layer.  

At a given position in the moir{\'e} pattern, the substrate interaction term in the moir{\'e} band 
Hamiltonian reflects the local coordination between the graphene sheet and the substrate lattice, 
{\em i.e.} the positions of nearby boron and nitrogen atoms in the substrate relative to the positions of 
the carbon atoms in the graphene sheet.  The substrate 
interaction term can be evaluated using {\em ab initio} methods\cite{jung_ab_2014,jung_origin_2015} 
by calculating the rigid displacement dependence of the band Hamiltonian of commensurate honeycomb structures.

In this paper we use a moir{\'e} band Hamiltonian for graphene on hBN derived in this way.  The moir{\'e} 
band Hamiltonian is able to account for the lattice mismatch of around 1.7 percent between graphene and hBN, and for 
strains in the graphene lattice and the substrate, and can be 
applied at any relative orientation between graphene sheet and substrate.
\cite{jung_origin_2015}   In a previous work we explained in detail how 
these effects combine with electron-electron interactions to control the size of the 
the gap which opens at the Fermi level of neutral graphene sheets, {\em i.e.} at the primary Dirac points in 
momentum space.\cite{jung_origin_2015}  In this paper we will focus on the secondary Dirac points, and 
on gaps at the Fermi level of graphene sheets with $\pm 4$ electrons per moir{\'e} period.
These features reflect the scattering of bare graphene sheet electrons off the periodic part of 
the substrate interaction Hamiltonian, whereas the neutral sheet gaps reflect mainly the spatial 
average of the substrate interaction Hamiltonian.  

The moir{\'e} band Hamiltonian can be written as a sum of bare Dirac ($H_{\rm D}$) and substrate interaction
($H_{{\rm M}}$) contributions:
\begin{equation}
H=H_{\rm D}+H_{{\rm M}}.
\end{equation}
For practical calculations, we express this Hamiltonian operator as a matrix in momentum space:
\begin{align}
\langle \vec{k}',s'\lvert H\rvert \vec{k},s\rangle = & \delta_{\vec{k},\vec{k}'}H_{\rm D}(\vec{k})+{}\nonumber\\
&{}+\sum_{\vec{G}}\langle s'\lvert H_{{\rm M},\vec{G}}\rvert s\rangle \; \delta_{\vec{k}',\vec{k}+\vec{G}}
\label{eqn:hamiltonian}
\end{align}
where $s$ and $s'$ are sublattice indices, $\vec{k}$ and $\vec{k}'$ are wavevectors, 
$H_{{\rm M},\vec{G}}$ is the Fourier transform  of $H_{{\rm M}}(\vec{r})$ over one period of the moir{\'e} pattern,
and $\vec{G}$ is a moir{\'e} pattern reciprocal lattice vector. 

Ref. \onlinecite{jung_origin_2015} discusses three versions of the 
moir{\'e} band Hamiltonian $H_{\rm M}$.   In the first version neither graphene nor boron nitride atomic positions were 
allowed to relax under the the influences of inter-layer forces. 
This version gives very small band gaps at the primary Dirac point and is not consistent with experimental results. 
The other choices are to let just the graphene lattice relax, or to let both the graphene and boron nitride lattices relax. 
These lead to larger primary Dirac point gaps that are more in line with experimentally observed results, suggesting that strains 
play an essential role in samples with large period moir{\'e} patterns. 
In this paper we will use the version of the moir{\'e} band Hamiltonian that accounts for strains 
in both graphene and in the substrate, although we expect only relatively small quantitative substrate 
strain effects at the secondary Dirac points.  

\section{Moir{\'e} band structure and density-of-states}

\begin{figure}
  \centering
  \includegraphics[width=\columnwidth]{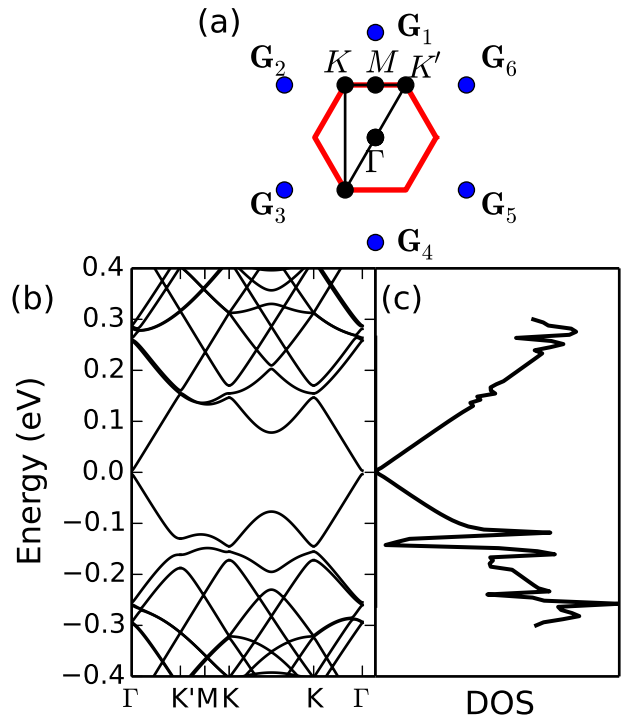}
  \caption{Band structure and density of states of graphene on boron nitride. (a) Schematic illustration of the moir{\'e} pattern
  Brillouin zone, outlined in red. 
The blue dots are moir{\'e} pattern reciprocal lattice vectors. 
We label high symmetry points in the moir{\'e} Brillouin zone, $\Gamma$, M, and K at 
the Brillouin zone center, edge center, and corner points respectively, by black dots.  
(b) Moir{\'e} bands along the black lines in (a). (c) 
The density of states (horizontal axis) as a function of energy (vertical axis).}
  \label{fig:bandstructure}
\end{figure}

\begin{figure}
  \centering
  \includegraphics[width=\columnwidth]{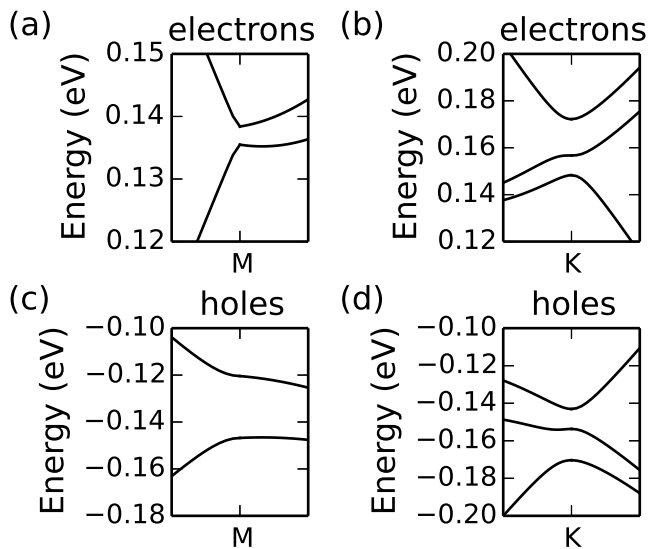}
  \caption{Band structure near the secondary Dirac points.  When substrate interactions are neglected, degeneracies
  forming secondary Dirac points occur at the M and K high-symmetry points defined in Figure~\ref{fig:bandstructure}.  
  The substrate interaction Hamiltonian lifts these degeneracies. There are 2-band avoided crossing near the M point 
  ((a) and (c)) and three band avoid crossings near the K point ((b) and (d)) in
  both the conduction band and the valence band.}
  \label{fig:bandstructurezoom}
\end{figure}

The moir{\'e} band structure and the density-of-states for the case of zero twist angle between graphene and substrate 
hexagonal lattices and a lattice constant mismatch of $-0.017$ are illustrated in Figure \ref{fig:bandstructure}. 
Note that in addition to the small gap at the Dirac point, there are avoided crossings at the high-symmetry 
Brillouin-zone boundary points M and K.  The electronic structure in this region of 
energy is highlighted in the Figure \ref{fig:bandstructurezoom}. 
Because distinct points on the Brillouin-zone boundary are connected by reciprocal lattice vectors,
the size of the avoided crossing gaps is directly related to elastic scattering of bare graphene states 
off the substrate interaction Hamiltonian associated with the moir{\'e} pattern.  

There is a distinct particle-hole asymmetry between the conduction and valence bands which 
is apparent in the density-of-states (Figure \ref{fig:bandstructure} (c)) and has been discussed previously in Refs.~\onlinecite{yankowitz_emergence_2012,wallbank_generic_2013,mucha-kruczynski_heterostructures_2013},
in terms of a phenomenological substrate interaction Hamiltonian in which the number of free parameters
has been minimized using symmetry considerations (the phenomenological substrate interaction
Hamiltonian is compared with the interaction Hamiltonian derived from {\em ab initio} theory used here 
in Appendix A)  and in Ref.~\onlinecite{moon_electronic_2014} in terms of an effective Hamiltonian derived from a tight binding model.

To understand the physical mechanism behind this asymmetry, we use a 
nearly-free Dirac-electron approximation by treating the substrate-interaction 
Hamiltonian as a perturbation at the M point.    
When reduced to the Brillouin-zone, two eigenstates of the bare graphene Dirac Hamiltonian,  
corresponding to Dirac cones centered at $\vec{G}_{0}\equiv 0$ and $\vec{G}_{4}\equiv G(0,-1)$ 
are degenerate at the M point. Here $G\equiv\lvert \vec{G}\rvert$ is the magnitude of the primitive moir{\'e} reciprocal lattice vectors. 
The perturbed energies are,
\begin{equation}
E^{(b)}_{\pm} = E_{0,{\rm M}}^{(b)}\pm\lvert U_{b}\rvert
\end{equation}
where $b$ is the band index ($1$ for conduction band and $-1$ for valence band), $E_{0,{\rm M}}^{(b)}=b \hbar v |G|/2$ is the energy of the
bare Dirac Hamiltonian states at the M point,
\begin{equation}
U_{b}=\left(\psi_{\vec{G}_{4}}^{(b)}\right)^{\dagger}H_{{\rm M},\vec{G}_{4}}\psi_{0}^{(b)}
\end{equation}
and $\psi_{\vec{G}}^{(b)}$ is the wavevector-dependent sub lattice spinor 
of the unperturbed Dirac Hamiltonian. This leading order perturbation theory analysis 
implies a band-dependent energy splitting at the M point equal to $\delta_{{\rm M},b}=2\lvert U_{b}\rvert$. 

To extract the physics behind the strong band dependence of this splitting, 
apparent in Figure \ref{fig:bandstructurezoom} and indirectly in Figure \ref{fig:bandstructure},
we decompose each Fourier component of the moir{\'e} band Hamiltonian into terms proportional to different sub lattice Pauli matrix contributions:\cite{kindermann_zero-energy_2012,yankowitz_emergence_2012,wallbank_generic_2013,mucha-kruczynski_heterostructures_2013,moon_electronic_2014}
\begin{equation}
H_{{\rm M},\vec{G}}=\sum_{\alpha=0,x,y,z}  h_{{\rm M},\vec{G}}^{\alpha} \; \tau^{\alpha}
\end{equation}
($\tau^{0}$ is the 2x2 identity matrix). Note that the expansion into Pauli matrices is justified by the property that $H_{{\rm M}}(\vec{r})$ is Hermitian; the non-Hermitian character of $H_{{\rm M},\vec{G}}$ allows the expansion coefficients $h_{{\rm M},\vec{G}}^{\alpha}$ to be complex.  
For the $\hat{y}$-direction M-point, the bare sub lattice spinor 
has a pseudospin orientation proportional to the $\hat{y}$-direction momentum.  It follows that,
\begin{align}
\left(\psi_{\vec{G}_{4}}^{(b)}\right)^{\dagger}\tau^{x}\psi_{0}^{(b)}\approx &b i\label{eqn:Mtaux} \\
\left(\psi_{\vec{G}_{4}}^{(b)}\right)^{\dagger}\tau^{y}\psi_{0}^{(b)} \approx &\; 0  \label{eqn:Mtauy} \\
\left(\psi_{\vec{G}_{4}}^{(b)}\right)^{\dagger}\tau^{z}\psi_{0}^{(b)}\approx & 1.0 \label{eqn:Mtauz},
\end{align}
and therefore that the coupling matrix element $U_{b}$ 
is produced by the $h_{{\rm M},\vec{G}}^{x}$ inter sub lattice tunneling and the $h_{{\rm M},\vec{G}}^{z}$ sub lattice site-energy
contributions to the substrate interaction  Hamiltonian.  The two sign choices 
in Eq.~\ref{eqn:Mtaux} correspond to conduction and valence bands.  From the substrate interaction Hamiltonian
\begin{align}
h_{{\rm M},\vec{G}_{4}}^{x}=&(-0.70-7.31i) \text{ meV}\approx -7.3i \text{ meV}\\
h_{{\rm M},\vec{G}_{4}}^{z}=&(-5.63-0.36i) \text{ meV}\approx -5.6 \text{ meV}
\end{align}
it follows that the two contributions to $U_{b}$ add in the valence band case and nearly cancel in the 
conduction band case.  The final result is that the gap at the M point in the moir{\'e} Brillouin-zone 
is $\delta_{{\rm M},v}=26$~meV in the valence band and almost an order of magnitude larger than in the conduction band where $\delta_{{\rm M},c}=2.9$~meV. 

\begin{figure*}
  \centering
  \includegraphics{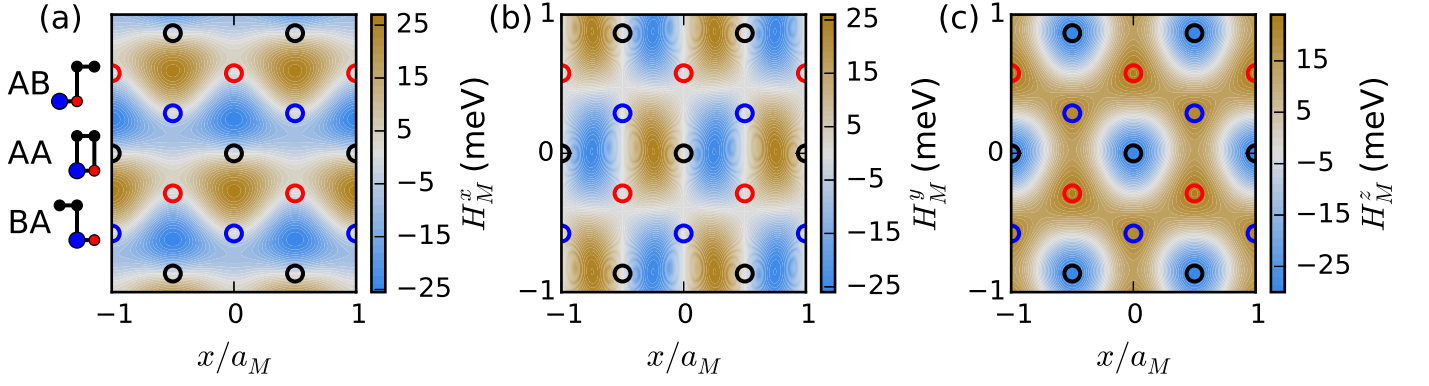}
  \caption{Real space maps of the moir{\'e}-band Hamiltonian. The moir{\'e} Hamiltonian is local in position, but sublattice-dependent.
At each position it can be decomposed into the sum of four terms proportional to Pauli matrices that act on sub lattice
degrees of freedom.  The $x$-component $H^{x}_{\rm M}$ is shown in (a), $H^{y}_{\rm M}$ in (b) and $H^{z}_{\rm M}$ in (c). 
For all subplots, the axes are in units of the moir{\'e} unit cell length, $a_{\rm M}$.  The open circles (color code explained by the  
cartoons in (a)) designate the positions within the moir{\'e} pattern of high-symmetry local stacking configurations.
Red open circles correspond to AB (carbon above nitrogen)  stacking, black to AA, and blue to BA (carbon above boron). 
In the cartoons, black dots are carbon atoms, reds are nitrogen, and blues are boron.
)}\label{fig:Hrealmaps}
\end{figure*}

Real space maps of the coefficients of the Pauli matrix expansion of the sub-lattice dependent 
substrate interaction Hamiltonian, $H^{x}_{\rm M}(\vec{r})$, $H^{y}_{\rm M}(\vec{r})$, and $H^{z}_{\rm M}(\vec{r})$, 
are provided in Figure \ref{fig:Hrealmaps}.  In our calculations the spatial origin is chosen to
lie at an AA point in Figure~\ref{fig:bandstructure}.  The relevant Fourier component for the matrix element we evaluate is $G_{4}$.
The weighting factor in $h_{{\rm M},\vec{G}_{4}}^{a}$  is therefore complex conjugated when $y \to -y$ in Figure~\ref{fig:Hrealmaps} where we 
see that $H^{z}_{\rm M}$ is approximately even and $H^{x}_{\rm M}$ is approximately odd.  
$H^z_{\rm M}$ provides a measure of the difference between site energies on the two graphene sub lattices.  
The difference between $\pi$-orbital energies is largest in magnitude when one carbon site is above 
the positively charged nitrogen site and the other is above the negatively charged boron site. 
At BA sites the carbon atoms above boron have a higher site energy than the the hexagonal 
plaquette centered carbon atoms, whereas at AB sites the plaquette centered carbon atoms have a higher site energy than the 
carbon atoms above nitrogen.  It follows that the on-site term $H^{z}_{\rm M}$ is negative at AA positions and positive at both AB and BA positions.
The pseudospin term $H^x_{\rm M}$ provides a measure of differences between the hopping amplitudes from a carbon 
atom to its three near neighbors, and this difference vanishes by symmetry at $AA$, $AB$ and $BA$ sites.
Carbon-carbon hopping is most anisotropic when one carbon atom is above the mid-point of a BN bond, and this 
leads to $H^x_{\rm M}$ values which have opposite sign at the mid-point between 
$AA$ and $AB$ points compared to the mid-point between $AA$ and $BA$ points,
and therefore to $h_{{\rm M},\vec{G}_{4}}^{x}$ values that are approximately imaginary.  
We see therefore that the particle hole asymmetry is described correctly only when both site energy and hopping amplitude distortions are accounted for properly.

At the $\hat{y}$-direction K point, there are three approximately degenerate bands, corresponding to
$\vec{G}_{0}=0$,  $\vec{G}_{4}= G(0,-1)$, and  $\vec{G}_{3}= G(-\sqrt{3}/2,-1/2)$. 
Perturbation theory (Appendix~\ref{sec:Kperturbation}) at K leads to a result that is similar to 
perturbation theory at M in that band separations are larger for the valence band than for
the conduction band.  Although the nearly-free-electron calculation is not as simple as in the 
two bands case, it is again true that a correct understanding of the 
origin of the strong particle-hole asymmetry requires an accurate account of substrate induced 
changes in both $\pi$-band energies and $\pi$-band hopping amplitudes.

\section{Transport theory}

With this background established, we now explore the {\em dc} transport properties of 
graphene on boron nitride.  One possible approach is to apply 
Boltzmann transport theory to the bands predicted by the 
moir{\'e} band Hamiltonian.  This strategy is however reliable
only when the associated Bloch state energy uncertainty, $\hbar/\tau$ where $\tau$ is the 
Bloch state lifetime, 
is small compared to the energy scale of band structure features.  
Because of the long period of the moir{\'e} pattern and the relatively weak strength of the substrate interaction,
the largest band structure feature scale is the 
$26$ meV valence band gap at four holes per moir{\'e} period explained in the previous section. 
We therefore choose an approach that is able to describe both weak and 
strong substrate interaction limits by applying 
a kinetic equation which is able to account for both 
intraband and interband contributions to the conductivity and using a relaxation time approximation for disorder.  
An overview of the theory is presented in this section, with details in Appendix~\ref{sec:transportTheory}. 
The conductivity can be decomposed into intraband and interband terms, arising from scattering within a single band and scattering between bands. 
The intraband contribution is proportional to the density of states and the interband contributions become important 
when the spacings between bands close to the Fermi energy is smaller than $\sim \hbar/ \tau$.
In particular, because of the small band separations at the Brillouin zone boundary, 
we anticipate a large interband contribution to the conductivity at four electrons per moir{\'e} period.

We obtain an estimate for the steady state density matrix by combining the equation of motion for the 
density-matrix ($\rho$) with a relaxation time approximation that
accounts for the influence of disorder scattering on both band diagonal and band off-diagonal terms: 
\begin{equation}\label{eqn:kinetic}
\frac{\partial\rho}{\partial t}=-\frac{i}{\hbar}\commutator{H}{\rho}+\frac{1}{\hbar}\frac{\partial\rho}{\partial\vec{k}}\cdot e\vec{E}-\frac{\rho-\rho_{0}}{\tau}
\end{equation}
Here, $H$ is the moir{\'e} band Hamiltonian in Equation~\eqref{eqn:hamiltonian}, $e$ is the magnitude of electric charge, $\vec{E}$ is the applied electric field, 
$\vec{k}$ is wave-vector in the moir{\'e} Brillouin zone measured from one of the graphene valleys, and $\tau$ is the relaxation time. 
The first term on the right-hand-side is purely off-diagonal in a moir{\'e} band Bloch representation and vanishes in 
equilibrium.  
The second term on the right-hand-side is the forcing term due to electric field, and the last 
term is the relaxation time approximation for scattering. 
We treat the relaxation time as an adjustable independent parameter, and assume that $\tau$ is the same for
both intraband and interband scattering. 
In the steady state, $\partial\rho/\partial t=0$. 
We expand this equation to linear order in the electric field $\vec{E}$, writing the density matrix, $\rho=\rho^{(0)}+\rho^{(1)}$, with 
$\rho^{(1)}$ being the linear response correction to the equilibrium moir{\'e} band density-matrix $\rho^{(0)}$:
\begin{equation}
\rho^{(0)} = \sum_{n\vec{k}}  \vert n\vec{k}\rangle \langle n\vec{k} \vert  \; f_{n\vec{k}}  
\label{eq:rho0}
\end{equation} 
where $n$ is a band index, and $f_{n\vec{k}}$ is one if the state is filled and zero if the state is empty. 
Since the first and third terms on the right-hand-side of Equation~\eqref{eqn:kinetic} are proportional to $\rho^{(1)}$,
the kinetic equation is readily solved for the density-matrix linear response.  The linear response current,
\begin{align}
j_{\alpha}=& \text{Tr }\{\rho^{(1)} \hat{j}_{\alpha}\} \equiv\sum_{n\vec{k}}\langle n\vec{k}\rvert\rho^{(1)} \hat{j}_{\alpha}\lvert n\vec{k}\rangle = \sum_{nm;\vec{k}}\rho_{nm}(-ev_{D}\tau^{\alpha})_{mn}\nonumber\label{eqn:current}\\
=& -ev_{D}\sum_{n\vec{k}}\left\{\rho^{(1)}_{nn}\langle n\vec{k}\rvert{\tau^{\alpha}}\lvert n\vec{k}\rangle+\sum_{m\neq n}\rho^{(1)}_{nm}\langle m\vec{k}\rvert{\tau^{\alpha}}\lvert n\vec{k}\rangle\right\}\\
\equiv & {j}_{\alpha,intra}+{j}_{\alpha,inter}
\end{align}
where $\alpha=x,y$ is the direction, $\hat{j}$ is the current operator and $v_{D}$ is the Dirac velocity of graphene.  
Note that $\tau$ without a superscript refers to relaxation time, while $\tau^{\alpha}$ with a superscript is a Pauli matrix. 
The conductivity can be correspondingly written as a sum of an intraband 
contribution due the dependence in Eq.(~\ref{eq:rho0}) of band eigenenergies on $\vec{k}$ and interband contributions due to the 
dependence of $\vec{k}$ variation of band eigenstates: 
\begin{align}
\sigma_{\rm intra}^{\alpha\beta}=&e^{2} v_{\rm D}^{2}\tau_{1}\sum_{n\vec{k}}\delta(\varepsilon_{\rm F}-\varepsilon_{n\vec{k}})\langle n\vec{k}\lvert\tau^{\alpha}\rvert n\vec{k}\rangle\langle n\vec{k}\lvert\tau^{\beta}\rvert n\vec{k}\rangle\\
\sigma_{\rm inter}^{\alpha\beta}=&ie^{2}\hbar v_{\rm D}^{2}\sum_{n,\vec{k},m\neq n} \frac{f_{m\vec{k}}-f_{n\vec{k}}}{\varepsilon_{n\vec{k}}-\varepsilon_{m\vec{k}}}\frac{\langle n\vec{k}\lvert \tau^{\alpha}\rvert m\vec{k}\rangle\langle m\vec{k}\lvert \tau^{\beta}\rvert n\vec{k}\rangle}{\left(\varepsilon_{n\vec{k}}-\varepsilon_{m\vec{k}}+i\hbar\tau_{2}^{-1}\right)}
\end{align}
where $\varepsilon_{n,\vec{k}}$ is the energy of band $n$ at wavevector $\vec{k}$ and $\tau_{1(2)}$ is the relaxation time for intra-(inter-)band scattering. 
The conductivity estimates summarized below were obtained by evaluating the sums in the above expressions numerically.  

To capture qualitative experimental features realistically, we used mobility as a parameter and from this calculated a relaxation 
time proportional to mobility and energy, which is appropriate for a linear band structure.\cite{das_sarma_electronic_2011}  
The relaxation time for band $b$ 
\begin{equation}
 \tau_{b} = \frac{\mu}{e}\frac{E}{v_{b}^{2}}
\end{equation}
is dependent on the mobility $\mu$, the energy $E$ measured from the primary Dirac point, and an approximate band velocity $v_{b}$. 
In the absence of a moir{\'e} pattern this expression for the relaxation time is motivated by the experimental finding that the conductivity in graphene sheets is proportional to carrier density - i.e. that although the mobility can be sample dependent, it is approximately independent of density in individual samples.  This property of graphene is related to the dependence of the disorder scattering amplitude on momentum transfer.\cite{ando_screening_2006,nomura_quantum_2007,adam_self-consistent_2007}  
We calculate the band velocity by taking the average of the velocity along the direction from the moir{\'e} zone center to the zone corner ($\Gamma$ to K in Fig.~\ref{fig:bandstructure}(a)), and the velocity along the direction from the zone center to the zone edge ($\Gamma$ to M in Fig.~\ref{fig:bandstructure}(a)).  We note that this procedure introduces a spurious reduction in the band velocity for the higher energy band of around 10 percent.  While both this velocity averaging and the isotropic assumption for $\tau$ are rough estimates, we expect this approximation will capture the main features of the electron transport.   
Table~\ref{table:tau} gives the average band velocities and representative relaxation times for the six lowest energy bands. 
For calculations of the mean free path and quantum transport for graphene on boron nitride without the effects of in-plane strain, we refer the reader to Ref.~\onlinecite{martinez-gordillo_transport_2014}.

\begin{table}
\caption{\label{table:tau}Table of band velocities and relaxation times for the six lowest energy bands of aligned graphene on boron nitride. 
The first two columns give the band edges, $E_{\rm min}$ and $E_{\rm max}$. 
Band velocity, $v_{b}$ is measured in units of the Dirac velocity for graphene, $v_{D}=0.84\times 10^{6}$ m/s. 
Both $\tau$ and $\hbar/\tau$ are given at an average band energy $E=(E_{\rm max}+E_{\rm min})/2$ for a sample mobility of $50,000$ cm$^{2}$/Vs.
Note that according to our calculations there is a finite energy gap in the valence band which is indirect in 
moir{\'e} momentum space and therefore smaller than the local valence band gaps at individual momenta.}
\begin{ruledtabular}
\begin{tabular}{lllll}
$E_{\rm min}$ (meV) & $E_{\rm max}$ (meV) & $v_{b}/v_{D}$ & $\tau_{b}$ (ps) & $\hbar/\tau_{b}$ (meV) \\
 -265 & -170  & 0.393 & 9.9 & 0.066 \\ 
 -253 & -147  & 0.752 & 2.5 & 0.26 \\
 -143 & -1.71 & 0.916 & 0.61 & 1.1 \\
 5.31 & 155 & 0.969 & 0.60 & 1.1 \\
  138 & 260 & 0.801 & 2.2 & 0.30 \\
  163 & 266 & 0.447 & 7.6 & 0.087 \\
\end{tabular}
\end{ruledtabular}
\end{table}

\begin{figure}
  \centering
  \includegraphics[width=\columnwidth]{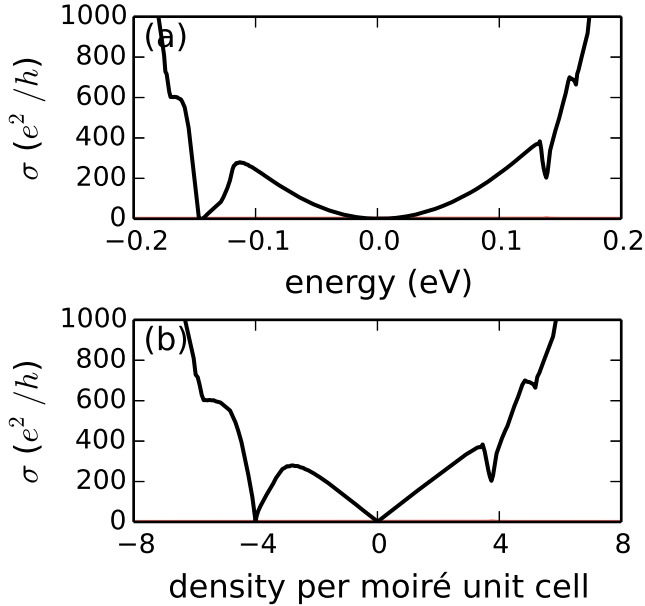}
  \caption{The conductivity of graphene on hBN as a function of (a) energy and (b) density for a mobility of $50,000$ cm$^{2}$/Vs. In both plots the total conductivity is a solid black line. Features in the density of states (see Figure~\ref{fig:bandstructure}~(c)) have corresponding features in the intraband contribution to conductivity. The interband conductivity (red lines) is negligible at this mobility.}
  \label{fig:totalconductivity}
\end{figure}

\section{Transport theory results}

\begin{figure}
  \centering
  \includegraphics[width=\columnwidth]{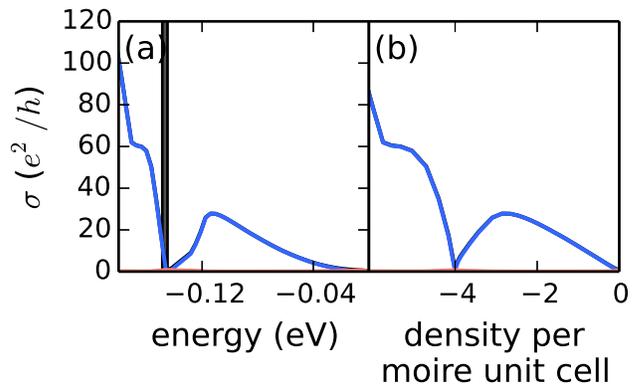}
  \caption{Conductivity {\em vs.} hole density. (a) The conductivity as a function of Fermi energy for p-type systems. The intraband contribution is 
  plotted in blue and the interband contribution in red. The gray shaded 
  region  indicates the energy range with the Fermi level in the gap at four holes 
  per moir{\'e} period.  (b) The conductivity as a function of hole density. The feature at four holes per moir{\'e} unit cell 
  is due to the gap at this density.  The calculations in this Figure are for a mobility of $5,000$ cm$^{2}$/Vs.}
  \label{fig:v1conductivity}
\end{figure}

\begin{figure}
  \centering
  \includegraphics[width=\columnwidth]{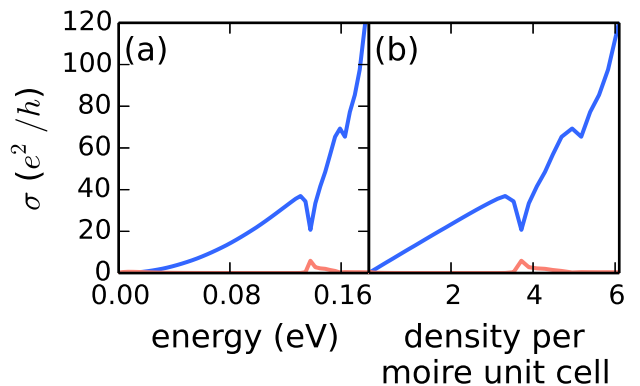}
  \caption{Conductivity {\em vs.} electron density. (a) The conductivity as a function of Fermi energy for n-type systems. The intraband contribution is 
  plotted in blue, the interband contribution in red. (b) The conductivity as a function of electron density. 
  The feature at four electrons per moir{\'e} unit cell is due to the avoided band crossings at the Brillouin-zone boundary which are not sufficiently strong to 
  yield an overall gap.  The calculations in this Figure are for a mobility of $5,000$ cm$^{2}$/Vs. The interband contribution has peaks when the interband separation (see Figure~\ref{fig:bandstructure}~(b)) is smallest}
    \label{fig:c1conductivity}
\end{figure}

The conductivity calculated for graphene on orientationally aligned hBN  is plotted in Figure \ref{fig:totalconductivity}  both as a 
function of Fermi energy and as a function of carrier density. At high mobilities, the total conductivity is indistinguishable from the intraband contribution. Its features closely track the density of states. 
On the other hand, at low mobilities the interband contribution peaks when the Fermi level is close to
weakly split Brillouin-zone edge states. 
Figures \ref{fig:v1conductivity} and \ref{fig:c1conductivity} show that while the interband contribution to the conductivity is generally quite weak, 
 there is a peak on the conduction band side at a density close to four electrons per moir{\'e}
period (red lines).  

Figure \ref{fig:v1conductivity} focuses on the strongest substrate related features in transport which 
appear when the Fermi level lies in the valence band at four holes per moir{\'e} period. 
Although much smaller, at $\delta_{v}\approx 3.5$ meV, than the splitting at individual k-points 
on the zone boundary, an overall gap does survive at this density.  
The gap is indirect in momentum space
with the valence band maximum at the moir{\'e} M point and the conduction band minimum 
at the moir{\'e} K point, as seen in Figure \ref{fig:bandstructurezoom} (c) and (d). When the mobility decreases $\hbar/\tau$ increases and the interband peak strengthens, 
slightly weakening the conductivity feature at $4$ holes per moir{\'e} period; 
$\hbar/\tau$ is $\sim0.1$~meV at this energy in Figure \ref{fig:totalconductivity} and $\hbar/\tau\sim1.0$~meV at this energy in Figure \ref{fig:v1conductivity}. In both cases, $\hbar/\tau$ is much smaller than the band splitting, and therefore the interband scattering is negligible on the hole side.

\begin{figure}
  \centering
  \includegraphics[width=\columnwidth]{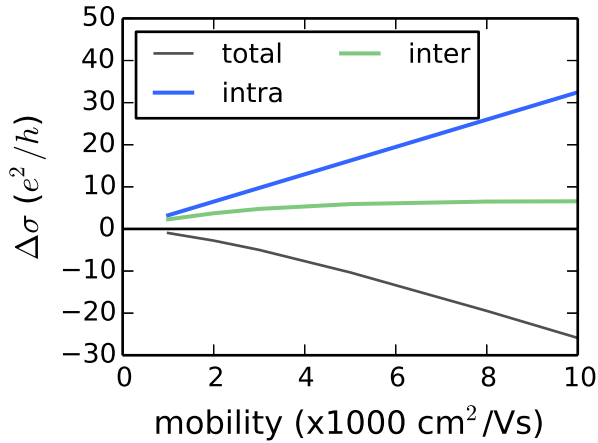}
  \caption{Mobility-dependence of the conductivity feature at four electrons per moir{\'e} period. 
  The size of the intraband conductivity (blue) dip and the interband conductivity peak (green) as a function of mobility. 
  The total feature size is shown in black.  For low mobility samples the relaxation approximation conductivity 
  can have a peak rather than a dip at this density.}
  \label{fig:conductivityvsmobility}
\end{figure}

A detailed look at the conduction band feature at a mobility of $5,000$~cm$^{2}$/Vs and four electrons per unit cell
is provided in Figure \ref{fig:c1conductivity}. The intraband conductivity (blue lines) shows a dip when 
the Fermi level is close to the energies of the moir{\'e} Brillouin zone boundary avoided crossing states.
There is no overall gap in the conduction band, as shown in Figure~\ref{fig:bandstructurezoom}, so there is a Fermi surface and an intraband 
contribution to transport at all energies in this interval, although the curve nevertheless has a dip. 
The peaks in the DOS are associated with saddle points in the moir{\'e} band structure. In particular, the first large peak is due to the saddle point at the M point in the first conduction band. The saddle point peaks are smoothed out in the conductivity calculation because of finite spectral widths associated with the finite Bloch state lifetimes. 
As shown in Figure~\ref{fig:c1conductivity}~(b), the dip in intraband conductivity is partially compensated 
by a peak in the interband contribution shown in red in Figure~\ref{fig:c1conductivity}.  
In Figure~\ref{fig:conductivityvsmobility} we plot the magnitude of both features as a 
function of mobility.   In the relaxation time approximation the intraband conductivity is proportional to mobility, and 
the intraband dip therefore dominates at high mobilities.  In the same approximation, the peak in the 
interband contribution can strongly compensate at low mobilities.  

Because the avoided crossing gaps on the moire band Brillouin-zone boundary are vastly larger in the hole carrier case than in the electron carrier case,
the physics of the conductivity minimum is not the same.  
In the valence band case, the size of the gap is large compared to $\hbar/\tau$ and the interband contribution to the dc conductivity is negligible.  
In the conduction band case, the density of states does not vanish at any energy so the intraband conductivity is always finite.  In addition avoided crossing gaps are not typically large compared to $\hbar/\tau$, allowing for a non-negligible interband contribution.  

\section{Summary and Discussion}

We have calculated the band structure, the density of states, and the 
transport properties of graphene on hexagonal boron nitride at zero twist angle using a moir{\'e} band model. 
All exhibit a pronounced particle-hole asymmetry, which we have traced to a correlation between 
spatial variations of the difference between honeycomb sub lattice site energies, and
spatial variations in intersublattice hopping amplitude properties. 
These variations are correlated because both are related to the charge difference
between boron and nitrogen sites in the substrate.   
The difference between nearest neighbor hopping amplitudes in carbon-above-boron and carbon-above nitrogen regions
plays a particularly essential role.  

We focus our transport calculations on the first features in conduction and valence bands, where there are avoided crossings of the secondary Dirac points, corresponding to four carriers per moir{\'e} unit cell. 
The particle-hole asymmetry is seen in the transport.  We find there is an overall
gap of $\sim 3.5$ meV in the valence band, and no gap in the conduction band. 
We have included effects of interband and intraband response in the conductivity, the latter becoming important at lower mobility. 
In the relaxation time approximation, the interband {\em dc} conductivity has peaks at 0 and $\pm 4$ electrons per moir{\'e} unit cell, which are not negligible for low mobility samples.

Experiments in graphene on boron nitride to date have focused on high-mobility samples, with mobilities 
as high as $275,000$ cm$^{2}$/Vs at low temperature.\cite{zomer_transfer_2011} 
Resistance peaks at 4 carriers per moir{\'e} unit cell and a distinct particle/hole asymmetry have been reported in Ref.~\onlinecite{hunt_massive_2013} in 
high-mobility ($100,000$ cm$^{2}$/Vs) samples and in Ref.~\onlinecite{yang_epitaxial_2013} for a sample of moderate mobility ($5,000$ cm$^{2}$/Vs) in agreement with our calculations. 
Although low mobility CVD graphene on SiO$_{2}$ and BN is available,\cite{gannett_boron_2011} transport measurements of low mobility rotationally aligned samples have not been reported. 
The interband conductivity peak arises from large interband matrix elements of the current operator. 
We expect that optical experiments can readily probe this scattering mechanism in high mobility samples.\cite{abergel_infrared_2013} 

\begin{acknowledgments}
The work in Austin was supported by the Department of Energy Division of Materials Sciences and Engineering under grant DE-FG03-02ER45958, 
and by the Welch Foundation under grant TBF1473. 
The work in Singapore was supported by the National Research Foundation of Singapore under its Fellowship programme (NRF-NRFF2012-01). 
SA and JJ thank Marcin Mucha-Kruczyski for several useful conversations comparing the phenomenological model developed in Refs.~\onlinecite{wallbank_generic_2013} and~\onlinecite{mucha-kruczynski_heterostructures_2013} with the ab initio theory presented here, and I. Yudhistira for discussions. 
We acknowledge the use of computational resources from the Texas Advanced Computing Center. 
\end{acknowledgments}

\appendix
\section{Comparison between phenomenological and ab-initio substrate interaction Hamiltonians}\label{sec:wallbankparams}
Refs.\onlinecite{wallbank_generic_2013,mucha-kruczynski_heterostructures_2013} 
formulate a symmetry-based phenomenological approach that can be 
used to construct effective Hamiltonians for graphene on substrates and leads to effective 
Hamiltonians of the form:  
\begin{widetext}
\begin{align}
H= \hbar v\vec{k}\cdot\vec{\sigma} + w_{0}\hbar vG\sigma_{0} + \tilde{w}_{3}\hbar vG\sigma_{3}\tau_{3} &+ u_{0} \hbar vGf_{1}(\vec{r})+u_{3}\hbar vGf_{2}(\vec{r})\sigma_{3}\tau_{3}+u_{1}\hbar v\left[\hat{z}\times\nabla f_{2}(\vec{r})\right]\cdot\vec{\sigma}\tau_{3}+u_{2}\hbar v\nabla f_{2}(\vec{r})\cdot\vec{\sigma}\tau_{3}\nonumber\\
& {}+\tilde{u}_{0}\hbar vGf_{2}(\vec{r})+\tilde{u}_{3}\hbar vGf_{1}(\vec{r})\sigma_{3}\tau_{3}+\tilde{u}_{1}\hbar v\left[\hat{z}\times\nabla f_{1}(\vec{r})\right]\cdot\vec{\sigma}\tau_{3}+\tilde{u}_{2}\hbar v\nabla f_{1}(\vec{r})\cdot\vec{\sigma}\tau_{3}\label{eqn:wallbankHamiltonian}
\end{align}
\end{widetext}
where
\begin{equation}
f_{1}(\vec{r})= \sum_{m=1}^{6}e^{i\vec{G}_{m}\cdot \vec{r}}
\end{equation} 
and 
\begin{equation}
f_{2}(\vec{r})= -i\sum_{m=1}^{6}(-1)^{m}e^{i\vec{G}_{m}\cdot\vec{r}}
\end{equation}
The vectors $\vec{G}_{m}$ are the first shell of moir{\'e} reciprocal lattice 
vectors labeled as in Figure~\ref{fig:bandstructure}.  
We rewrite our effective Hamiltonian obtained by performing {\em ab initio} calculations 
in this phenomenological form.\cite{marcin_mucha_kruczynski_private}  In Table ~\ref{table:wallbankparams}  we list the Hamiltonian parameters 
we obtain in this way for a lattice mismatch of $\varepsilon=-0.017$ and a zero twist angle. 
We note that including relaxation in the Hamiltonian significantly changes the values of the parameters. 
These differences in parameters are relevant for the 
interpretation of the particle-hole asymmetry in graphene on hBN.  

\begin{table}
\caption{\label{table:wallbankparams}{Table of parameters for the Hamiltonian in Eqn.~\eqref{eqn:wallbankHamiltonian} at $\varepsilon = -0.017$ and zero twist angle. The first column shows parameters for the Hamiltonian used in this paper. This Hamiltonian includes relaxation effects of the graphene and boron nitride lattices. The second column shows the parameters without including relaxation effects. Both these Hamiltonians are discussed in detail in Ref.~\onlinecite{jung_origin_2015}. All quantities are in meV units.}}
\begin{ruledtabular}
\begin{tabular}{c d d}
 & \text{Relaxed} & \text{Rigid} \\ 
$w_{0}\hbar vG$ (meV)         & 0     & 0     \\
$\tilde{w}_{3}\hbar vG$  (meV)& 3.74  & 0     \\
$u_{0}\hbar vG$ (meV)         & 1.26  & -0.64\\
$\tilde{u}_{0}\hbar vG$  (meV)& 8.98  & 10.10\\
$u_{1}\hbar vG$ (meV)         & 0.70  & 1.97 \\
$\tilde{u}_{1}\hbar vG$  (meV)& -7.31 &-11.17\\
$u_{2}\hbar vG$  (meV)        & 0     & 0    \\
$\tilde{u}_{2}\hbar vG$  (meV)& 0     & 0    \\
$u_{3}\hbar vG$ (meV)         & -0.36 & 1.26 \\
$\tilde{u}_{3}\hbar vG$  (meV)& -5.63 & -8.89\\
\end{tabular}
\end{ruledtabular}
\end{table}

\section{Avoided crossing analysis at the moir{\'e} Brillouin zone corners}\label{sec:Kperturbation}
At each of the moir{\'e} Brillouin zone corners, there are three degenerate solutions to the gapped Dirac equation, $H_{D}+H_{{\rm M},\vec{G}=0}$ with a mass of $m=3.7$ meV.\cite{moon_electronic_2014,wallbank_generic_2013}   
As in the case of the M point shown in the main text, we treat the off-diagonal terms in a degenerate perturbation theory. 
The energies are the eigenvalues of the $3\times 3$ effective Hamiltonian,
\begin{equation}
H_{K,eff}^{(b)}=\left(\begin{array}{ccc}
  E_{0,K}^{(b)} & t_{3,b}^{*} & t_{4,b}^{*}\\
  t_{3,b} & E_{0,K}^{(b)} & t_{5,b}^{*} \\
  t_{4,b} & t_{5,b} & E_{0,K}^{(b)}\end{array}\right)
\end{equation}
where the index $\pm$ refers the conduction ($+$) or valence ($-$)  band and
\begin{equation}
t_{j,b} = \left(\psi_{\vec{G}_{i}+\vec{G}_{j}}^{(b)}\right)^{\dagger}H_{{\rm M},\vec{G}_{j}}\psi_{\vec{G}_{i}}^{(b)}
\end{equation}
The unperturbed energy on the diagonal is,
\begin{equation}
E_{0,K}^{(b)}=b\sqrt{m^{2}+(\hbar v G/\sqrt{3})^{2}}
\end{equation}
There are three eigenvalues of the Hamiltonian, labeled with subscript $i=1,2,3$ 
\begin{equation}
E_{i}^{(b)}=E_{0,K}^{(b)}-w_{i}-\frac{\lvert t_{3,b}\rvert^{2}+\lvert t_{4,b}\rvert^{2}+\lvert t_{5,b}\rvert^{2}}{3w_{i}}
\end{equation}
where $w_{i}$ are the three solutions to the polynomial
\begin{equation}
w^{6}+2\text{Re }\{t_{3}^{*}t_{5}^{*}t_{4}\}w^{3}+\left(\frac{\lvert t_{3,b}\rvert^{2}+\lvert t_{4,b}\rvert^{2}+\lvert t_{5,b}\rvert^{2}}{3}\right)^{3}=0
\end{equation}
which are guaranteed to be real since the Hamiltonian is Hermitian. 
Decomposing the moir{\'e} Hamiltonian into terms proportional to Pauli matrices,
\begin{equation}
H_{{\rm M},\vec{G}}=\sum_{\alpha=0,x,y,z}h_{M,\vec{G}}^{\alpha}\tau^{\alpha}
\end{equation}
($\tau^{0}$ is the 2x2 identity matrix), we 
obtain values for $h_{M,\vec{G}}$ and $\left(\psi_{\vec{G}_{i}+\vec{G}_{j}}^{(b)}\right)^{\dagger}\tau^{\alpha}\psi_{\vec{G}_{i}}^{(b)}$ shown in table \ref{table:Kcrossing}. The situation is much more complicated than that of the M point, but the generic features remain: there is a sign change in the matrix elements of $\tau^{x,y}$ and not $\tau^{0,z}$, or vice versa, which results in constructive interference of 
different contributions to the $t_{i,b}$ values in the valence band and destructive interference in the conduction band.
This leads to a smaller energy splittings in the conduction band.

\begin{table*}
\caption{\label{table:Kcrossing}Parameters for the analysis of the avoided crossing at the K point. The first three rows are the Pauli matrix decompositions of the moir{\'e} potential for values of reciprocal lattice vector relevant to the perturbation theory at the K point in the moir{\'e} Brillouin zone. The rest of the table shows the matrix elements of the Pauli matrices with the unperturbed wavefunctions at the given $\vec{G}$ values and $\vec{k}$ at the K point of the moir{\'e} Brillouin zone. Both conduction and valence band values are shown.}
\begin{ruledtabular}
\begin{tabular}{c c c c c}
$\alpha$               & $0$ & $x$ & $y$ & $z$ \\ \hline
$h^{\alpha}_{M,\vec{G}_{3}}$ (meV) & $1.26-8.98i$ & $0.350-3.66i$ & $-0.606+6.33i$ & $-5.63+0.36i$ \\
$h^{\alpha}_{M,\vec{G}_{4}}$ (meV) & $1.26+8.98i$ & $-0.699-7.31i$ & $0$ & $-5.63-0.36i$ \\
$h^{\alpha}_{M,\vec{G}_{5}}$ (meV) & $1.26-8.98i$ & $0.350-3.66i$ & $0.606-6.33i$ & $-5.63+0.364i$ \\ \hline
Conduction band:       &   &   &   &   \\
$\left(\psi_{\vec{G}_{3}}^{(+)}\right)^{\dagger}\tau^{\alpha}\psi_{\vec{G}_{0}}^{(+)}$ 
    & $0.268-0.423i$ & $-0.250+0.433i$ & $0.433-0.750i$ & $0.756+0.423i$ \\ 
$\left(\psi_{\vec{G}_{4}}^{(+)}\right)^{\dagger}\tau^{\alpha}\psi_{\vec{G}_{0}}^{(+)}$ 
    & $0.268+0.423i$ & $0.500+0.866i$ & $0$ & $0.756-0.423i$ \\ 
$\left(\psi_{\vec{G}_{4}}^{(+)}\right)^{\dagger}\tau^{\alpha}\psi_{\vec{G}_{3}}^{(+)}$ 
    & $0.268-0.423i$ & $-0.250+0.433i$ & $-0.433+0.750i$ & $0.756+0.423i$ \\ \hline
Valence band:          &   &   &   &   \\
$\left(\psi_{\vec{G}_{3}}^{(-)}\right)^{\dagger}\tau^{\alpha}\psi_{\vec{G}_{0}}^{(-)}$ 
    & $-0.232+0.443i$ & $-0.250+0.433i$ & $0.433-0.750i$ & $-0.744-0.443i$ \\ 
$\left(\psi_{\vec{G}_{4}}^{(-)}\right)^{\dagger}\tau^{\alpha}\psi_{\vec{G}_{0}}^{(-)}$ 
    & $-0.232-0.443i$ & $0.500+0.866i$ & $0$ & $-0.744+0.443i$ \\ 
$\left(\psi_{\vec{G}_{4}}^{(-)}\right)^{\dagger}\tau^{\alpha}\psi_{\vec{G}_{3}}^{(-)}$ 
    & $0.232-0.443i$ & $0.250-0.433i$ & $0.433-0.750i$ & $0.744+0.443i$ \\
\end{tabular}
\end{ruledtabular}
\end{table*}

\section{Transport theory}\label{sec:transportTheory}
The commutator, $\left[H,\:\rho\right]$, is
\begin{equation}
\left[H,\:\rho\right]_{nn'}=\left(\varepsilon_{nk}-\varepsilon_{n'k}\right)\rho_{nn'}
\end{equation}
which vanishes when $n=n'$.

The derivative with respect to $k$ of the equilibrium density matrix is,
\begin{align}
\frac{\partial\rho^{(0)}}{\partial\vec{k}}=& \sum_{m}\left\{\frac{\partial f_{m\vec{k}}}{\partial\vec{k}}\lvert mk\rangle\langle mk\rvert+{}\right.\nonumber\\
& \left.{}+f_{m\vec{k}}\left\lvert\frac{\partial}{\partial\vec{k}}\: mk\right\rangle\langle mk\rvert+f_{m\vec{k}}\lvert mk\rangle\left\langle \frac{\partial}{\partial\vec{k}}\: mk\right\rvert\right\}
\end{align}
The derivative of the wavefunction with respect to k is,
\begin{equation}
\left\lvert\frac{\partial}{\partial\vec{k}}\: nk\right\rangle = \sum_{m\neq n}\left(\frac{\langle mk\rvert\frac{\partial H}{\partial \vec{k}}\lvert nk\rangle}{\varepsilon_{nk}-\varepsilon_{mk}}\right)\lvert mk\rangle
\end{equation}
The only part of $H$ which depends explicitly on $\vec{k}$ is $H_{0}=\hbar v_{\rm D}\left(\tau^{x},\,\tau^{y}\right)\cdot\vec{k}$ which has derivative with respect to $k_{\alpha}$ of $\hbar v_{\rm D}{\tau^{\alpha}}$.
It follows that 
\begin{align}
\left(\frac{\partial\rho^{(0)}}{\partial{k_{\alpha}}}\right)_{nn'}&=\delta_{nn'}\frac{\partial f}{\partial\varepsilon_{nk}}\langle nk\rvert \hbar v_{\rm D}{\tau^{\alpha}}\lvert nk\rangle +{}\nonumber\\
&{}+(1-\delta_{nn'})\langle nk\rvert \hbar v_{\rm D}{\tau^{\alpha}}\lvert n'k\rangle\frac{f_{n'\vec{k}}-f_{n\vec{k}}}{\varepsilon_{n'k}-\varepsilon_{nk}}
\end{align}
Note that $\tau$ without a superscript refers to relaxation time, while $\tau^{\alpha}$ with a superscript is a Pauli matrix. Using this in Equation~\eqref{eqn:kinetic} gives the expression for the current, Equation~\eqref{eqn:current}.

%\bibliography{zotero}


\end{document}